%% file: lambda.tex
\documentclass[12pt,a4paper,dvips]{article}
\usepackage{a4p} \usepackage{cite,mcite} \usepackage{graphicx}
\usepackage{physics} \usepackage{l3_title,ifthen,Lep}
\usepackage{epsfig,float}
\journalname{Phys. Lett. B} \preprint{2002-013} \date{February 6, 2002} \Lep{2}
\def\ra{\rightarrow } \def\epem{\mbox{e}^+\mbox{e}^- }   \def\la{\mbox{\Lambda} } \def\kos{\mbox{K}^0_s }
 \def\la{\Lambda }
\def\ssbar{\overline{\Sigma}\,\!^0} \def\llbar{\overline{\Lambda}}
\def\sisibar{\Sigma^0\overline{\Sigma}\,\!^0}
\def\lalabar{\Lambda\overline{\Lambda}} \def\kos{\rm \mbox{K}^0_s }
\def\koskos{\kos\kos} 

\newlength{\capindent} 
\newlength{\capwidth} 
 \newlength{\figwidth}
\newcommand{\icaption}[2][!*!,!]{\hspace*{\capindent}%
  \begin{minipage}{\capwidth}
    \ifthenelse{\equal{#1}{!*!,!}}%
      {\caption{#2}}%
      {\caption[#1]{#2}}
  \end{minipage}}
\begin{document}

\begin{titlepage}

\title{ $\mathbf{\Lambda}$ and $\mathbf{\Sigma^0}$ Pair Production \\
in Two-Photon Collisions at LEP }

\author{The L3 Collaboration}

\begin{abstract}
Strange baryon pair production in two-photon collisions is studied
with the L3 detector at LEP. The analysis is based on data collected
at $\epem$ centre-of-mass energies from 91 GeV to 208 GeV,
corresponding to an integrated luminosity of 844 pb$^{-1}$. The
processes $\gamma \gamma \ra \Lambda \llbar$ and $\gamma \gamma \ra
\Sigma^0 \ssbar$ are identified. Their cross sections as a function of
the $\gamma \gamma$ centre-of-mass energy are measured and results are
compared to predictions of the quark-diquark model.
\end{abstract}

\submitted

\end{titlepage}

\section{Introduction}

Electron-positron colliders are a suitable place for the study of
two-photon interactions, via the process $\epem \ra \epem \gamma^*
\gamma^* \ra \epem X$, where $\gamma^*$ denotes a virtual photon. The
outgoing electron and positron carry almost the full beam energy and
are usually undetected, due to their small transverse momenta. The
final state $X$ has, therefore, a low mass relative to the $\epem$
centre-of-mass energy, $ \sqrt{s}$. The small photon virtuality allows
the extraction of the cross section \mbox{$\sigma(\gamma \gamma \ra
X)$} in real photon collisions, once the photon flux is calculated by
QED \cite{b1}.

  The process \mbox{$\gamma \gamma \ra $ \sl baryon antibaryon} is
sensitive to the quark structure of the baryon. Calculations of the
cross section for this process were performed using the hard
scattering approach \cite{b15}. Due to the failure of a three-quark
calculation \cite{b16} to correctly predict the $\rm \gamma \gamma \ra
p \overline{p}\, $ cross section in the GeV region \cite{b17}, an
alternative quark-diquark model was proposed \cite{ans}. This model
includes non-perturbative effects through the use of diquarks, a qq
bound state within the baryon \cite{b3}.

In this letter we present the first measurements at LEP of the cross
sections $\sigma(\gamma \gamma \ra \Lambda \llbar)$ and $\sigma(\gamma
\gamma \ra \Sigma^0 \ssbar)$. We analysed a total integrated
luminosity of 844 pb$^{-1}$ collected with the L3 detector
\cite{b4}. Out of this sample, 157 pb$^{-1}$ were collected around the
Z peak and 687 pb$^{-1}$ at centre-of-mass energies from 161 GeV to
208 GeV. The analysis is based on the central tracking system and the
high resolution BGO electromagnetic calorimeter. The events are
selected by the track triggers \cite{b6}.

Monte Carlo events are generated \cite{b7} for each beam energy and
for each process within the formalism of Reference \citen{b1}. An
uniform spectrum as a function of the two-photon mass, $W_{\gamma
\gamma}$, from threshold to 5 GeV is used. The two-body final states
$\Lambda \llbar$ and $\Sigma^0 \ssbar$ are generated isotropically in
the centre-of-mass of the $\gamma\gamma$ system.  The events are then
passed through the full L3 detector simulation using the GEANT
\cite{b8} and GEISHA \cite{b9} programs and are reconstructed
following the same procedure as for the data. Time dependent detector
inefficiencies, as monitored during the data taking period, are taken
into account.

The CLEO, TPC/2$\gamma$ and VENUS collaborations \cite{b2,b19,b20}
searched for the reaction $\epem \ra \epem \Lambda \llbar$ at
$\sqrt{s}=10.6$ GeV, 14.5 GeV and 58 GeV, respectively. Only CLEO and
TPC/2$\gamma$ observe a signal. No results for the $\epem \ra \epem
\Sigma^0 \ssbar$ cross section were reported so far. Our results are
compared to these experiments and to theoretical predictions of the
quark-diquark model.

\section{Event selection}

  In order to study the $\lalabar$ and $\sisibar$ final states, the
  $\Sigma^0 \ra \Lambda \gamma$, $\ssbar \ra \llbar \gamma$, $\Lambda
  \ra {\rm p} \pi^-$ and $\llbar \ra {\rm \overline{p}} \pi^+$ decays
  are considered. The preselection of events is based on charged
  tracks and proceeds as follows:

\begin{itemize}

\item There must be four charged tracks in the tracking chamber with a
net charge of zero. These tracks must be reconstructed from at least
12 hits out of a maximum of 62.

\item There must be two secondary vertices at a distance from the
primary interaction vertex greater than $\rm 3 \,mm$ in the transverse
plane.

\item The angle in the transverse plane between the flight direction
of each $\Lambda$ candidate and the total momentum vector of the two
outgoing tracks must be less than 0.15 rad.

\item Events with a secondary vertex due to a photon conversion are
rejected. A conversion is identified if by assigning the electron mass
to the two tracks, their effective mass is below $0.05$ GeV.

\end{itemize}

\subsection{$\mathbf{\Lambda}$ identification}

\begin{figure}
 \begin{center}
  \epsfig{file=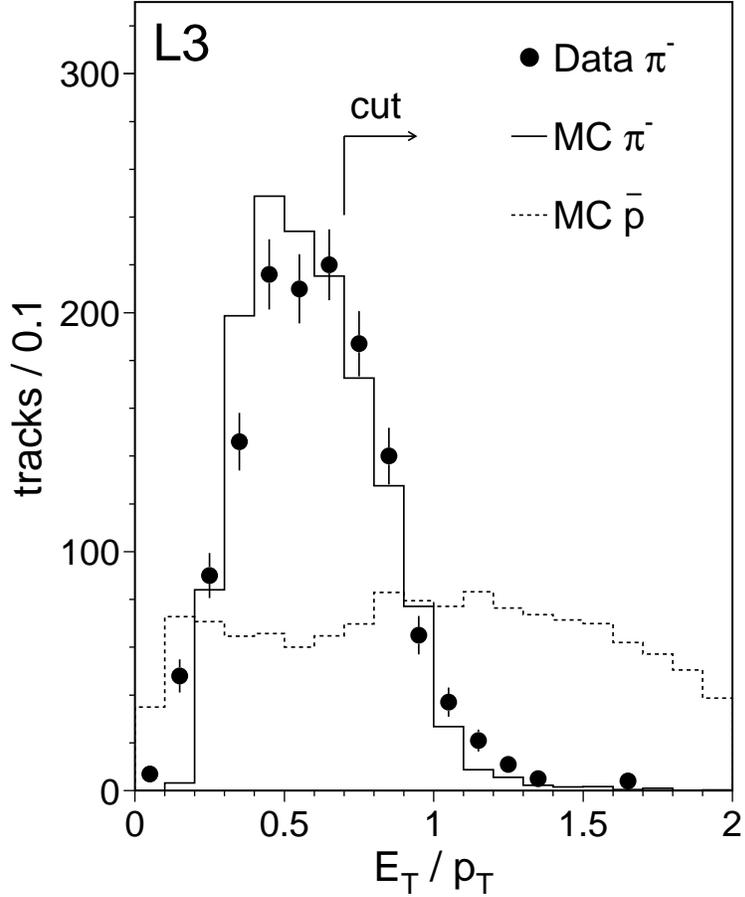,width=10cm}
 \end{center}
\caption {Distribution of the ratio of the transverse energy deposited in the 
electromagnetic calorimeter, $E_T$, and the transverse momentum,
$p_T$, for pions and antiprotons. The $\pi^-$ data distribution is
obtained from a high purity $\koskos$ sample~{\protect
\cite{b12}}. The $\pi^-$ Monte Carlo distribution corresponds to a
simulated $\ee\ra\ee \koskos$ sample normalized to the number of
$\koskos$ observed in data~{\protect \cite{b12}}. The antiproton
distribution is obtained from simulated $\ee\ra\ee\lalabar$ events,
with arbitrary normalization.}
\label{nrj}
\end{figure}

 The two secondary vertices are assigned to the $\Lambda$ and
 $\overline{\Lambda}$ decays. At each vertex, the p or the $\rm
 \overline{p}$ are identified as the highest momentum tracks. Monte
 Carlo studies show that this is the correct configuration in more
 than 99\% of the cases. To suppress the dominant $\epem\ra \epem
 \koskos \ra \epem \pi^+ \pi^- \pi^+ \pi^-$ background, the following
 criteria are applied:

\begin{itemize} 

\item The $dE/dx$ measurement must be consistent with the $\Lambda$ or
$\llbar$ hypotheses. A confidence level $\rm CL > 0.001$ is required
for both proton, antiproton and pions candidates.  This cut rejects
85\% of the $\koskos$ background.

\item The particle assignment is considered to be correct if either,

a) there is at least one of the tracks associated to a proton or to an
   antiproton with more than 30 hits and a $dE/dx$ confidence level
   ratio $\rm CL(p)/CL(\pi)>10$, or

b) the ratio between the electromagnetic transverse energy, $E_T$, and
   the transverse momentum, $p_T$, of the antiproton candidate is
   greater than 0.7. This cut eliminates 70\% of the pions and keeps
   77\% of the antiproton signal, as shown in Figure~\ref{nrj}.

The $dE/dx$ identification has a high discriminating power for
particles with momentum below 700 MeV, whereas the $E_T/p_T$ cut
becomes more efficient for higher momentum antiprotons. These criteria
suppress 83\% of the remaining $\koskos$ background.

\item If two $\kos$ candidates are reconstructed, the event is
rejected.  A $\kos$ candidate is defined as a system with a
reconstructed $\pi^+\pi^-$ mass within a $\pm \, 30$ MeV interval
around the nominal $\kos$ mass. Only 1\% of the original $\koskos$
background remains after this cut.

\end{itemize}

\begin{figure}
 \begin{center}
 \begin{tabular}{rl}
   \epsfig{file=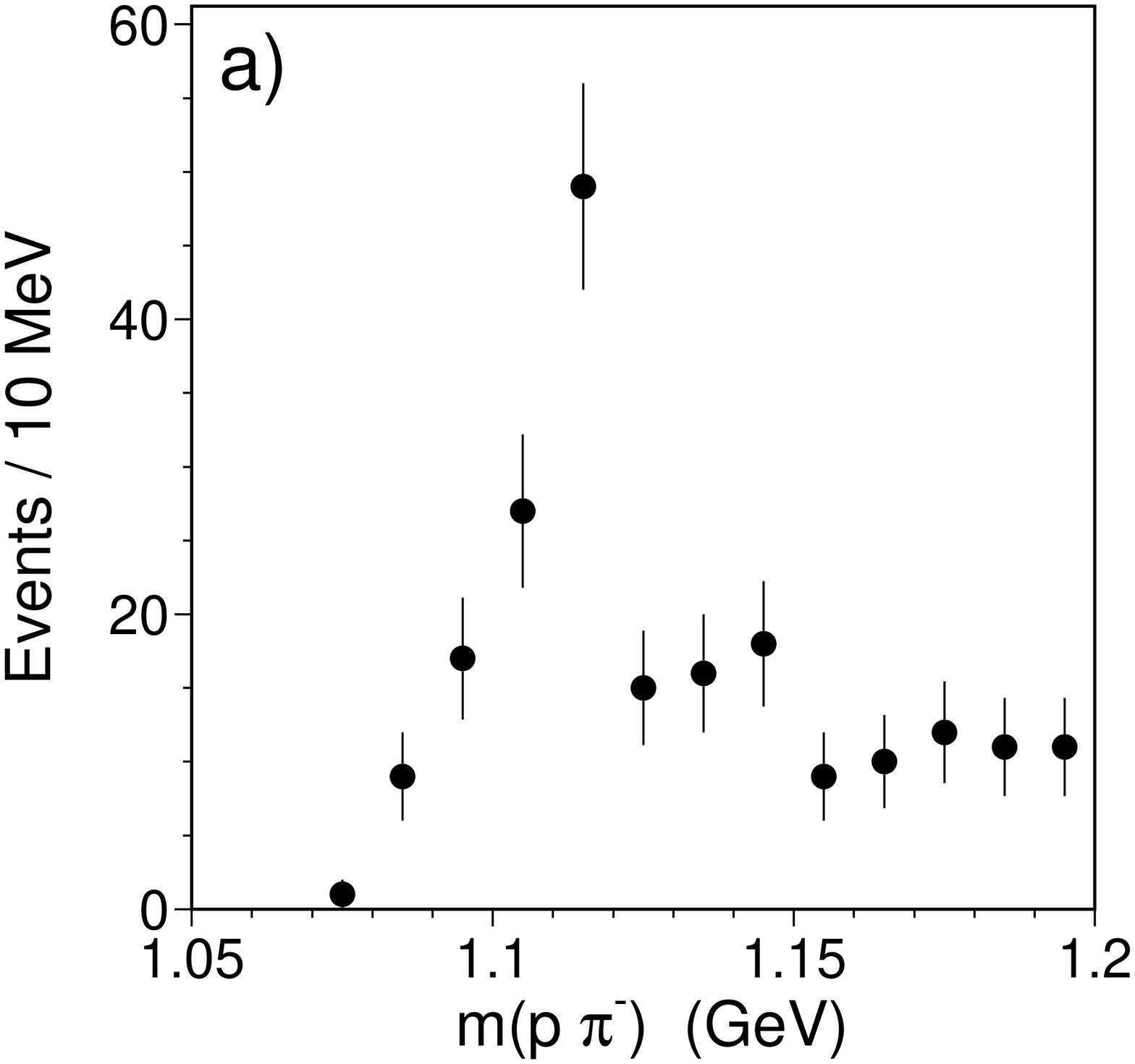,width=7.5cm} &
   \epsfig{file=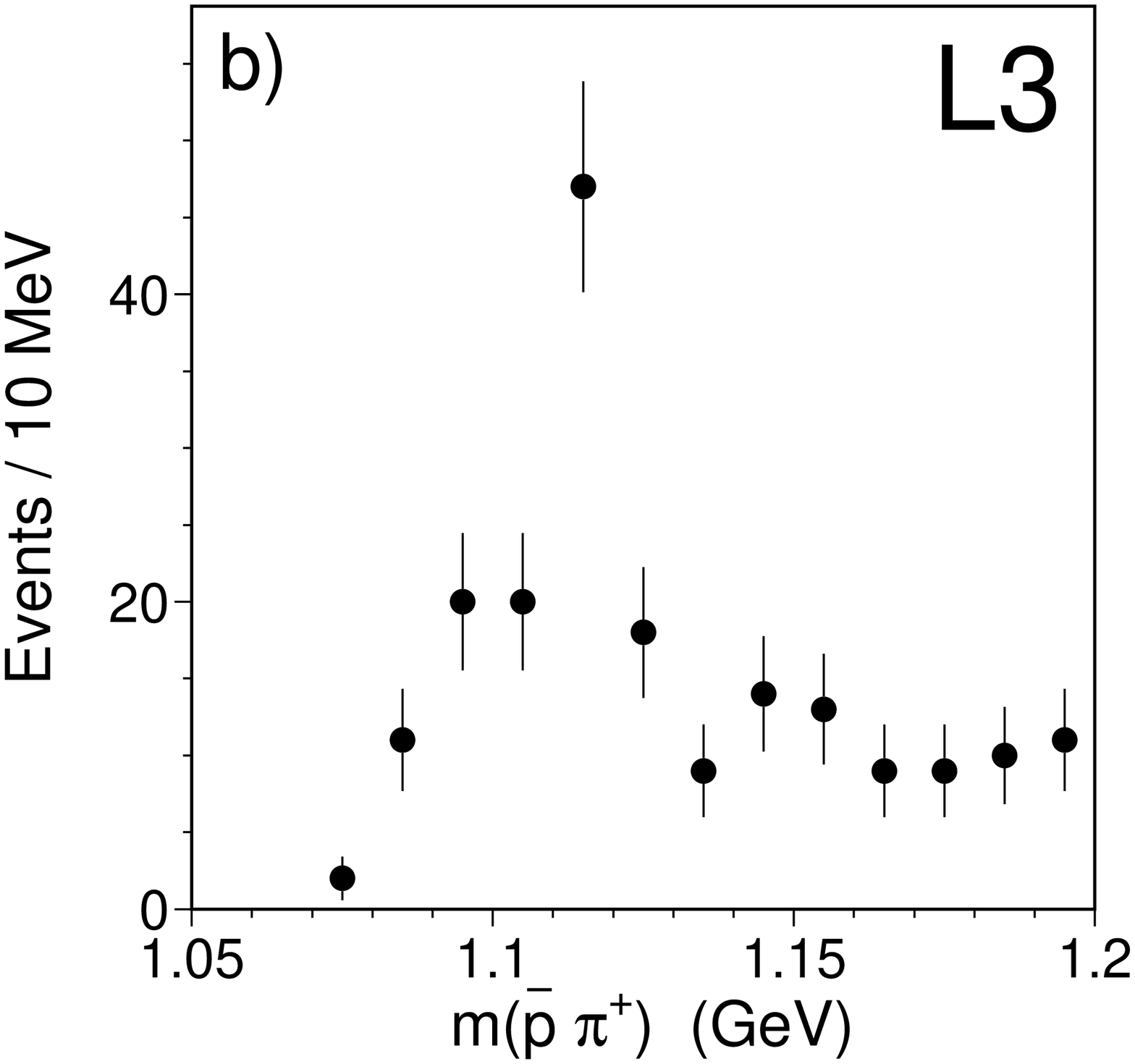,width=7.5cm} \vspace{0.5cm}\\
   \epsfig{file=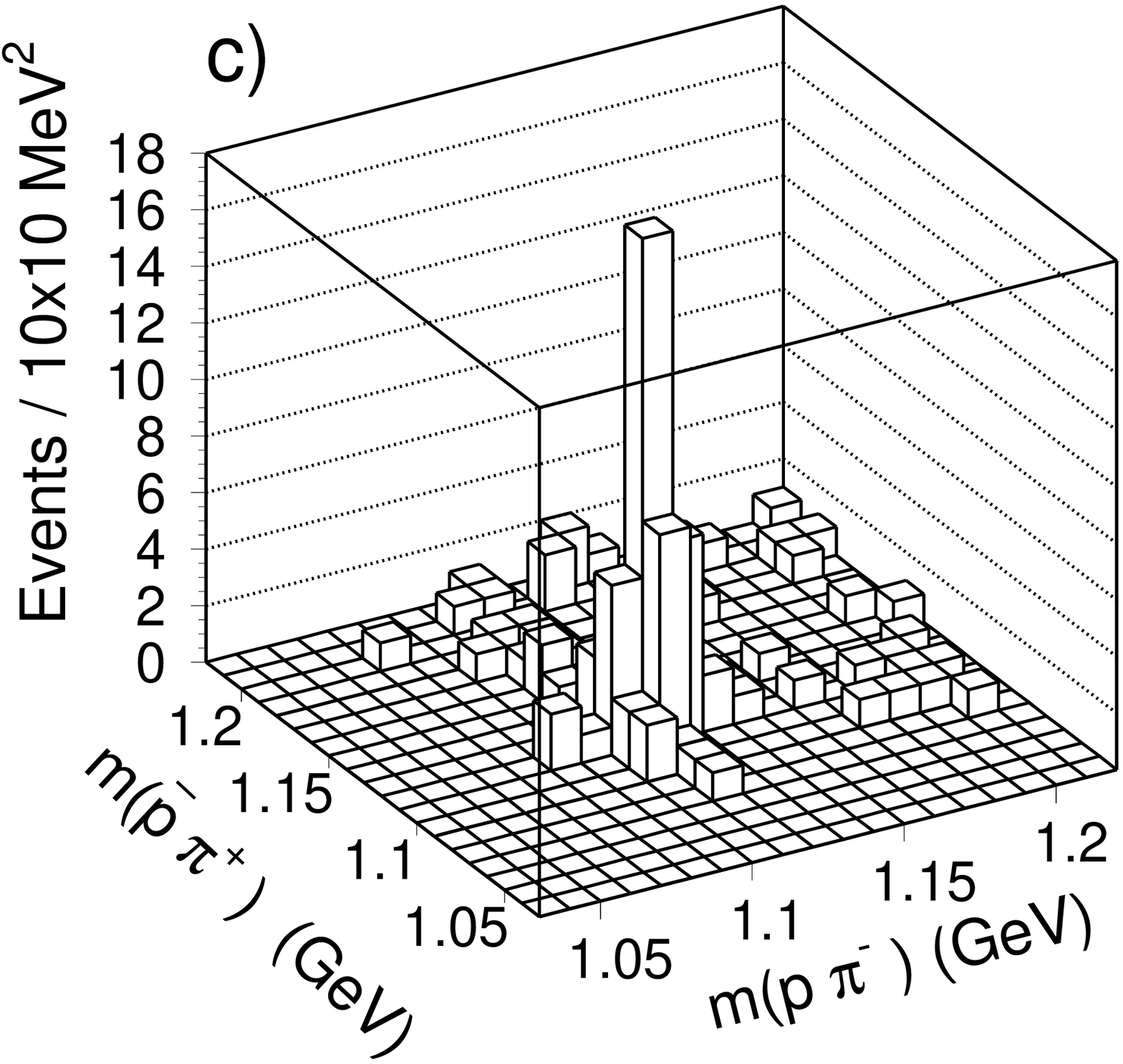,width=7.5cm} &
   \epsfig{file=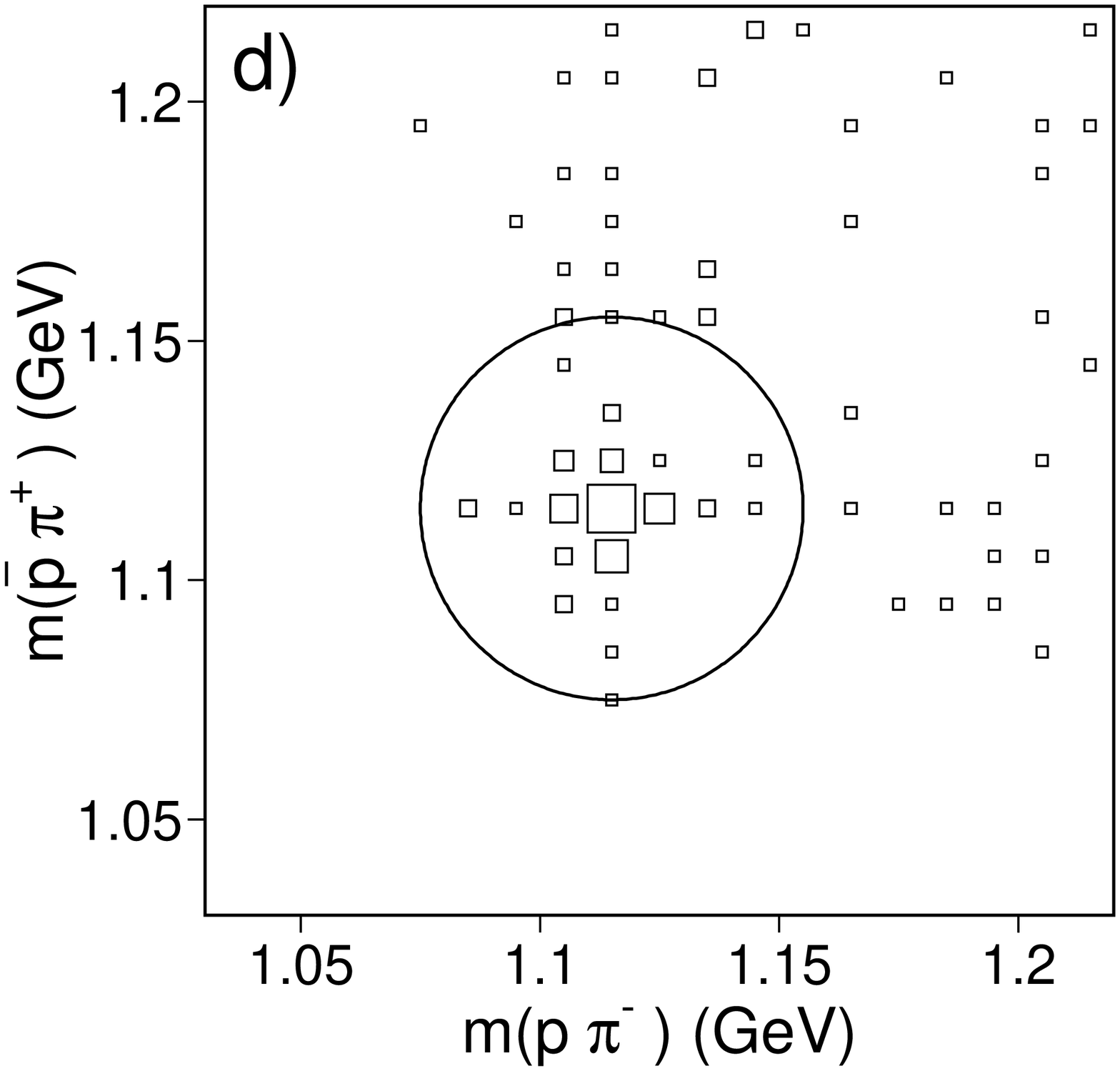,width=7.5cm}
 \end{tabular}
 \end{center} 
 \caption{Effective mass distribution of the a) $\rm p \pi^-$ system and
b) $ \overline{p} \pi^+$ system. The two dimensional distribution is
shown in c) and d). A radius of 40 MeV around the nominal mass value
of $m_{\Lambda}$ defines the inclusive $\lalabar$ sample.}
\label{mass}
\end{figure}

  In addition to the previous requirements, a cut $| \cos \theta^* | <
0.6$ is applied, where $\theta^*$ is the polar angle of the $\Lambda$
in the two-photon centre-of-mass system, to match the experimental
acceptance with the range of the theoretical predictions. Clean
$\Lambda$ and $\llbar$ signals are observed in the distributions of
the $\rm p\pi^-$ and $\rm \overline{p}\pi^+$ masses, presented in
Figures~\ref{mass}a and~\ref{mass}b. The $\Lambda$ and $\llbar$ masses
are found to be $m_{\Lambda}= 1.113 \pm 0.006$ GeV and $m_{\llbar}=
1.115 \pm 0.006$ GeV, respectively, in agreement with the nominal
value of 1.116 GeV \cite{b11}. The final sample contains 66 inclusive
$\lalabar$ candidates. They are selected within a radius of 40 MeV
around the nominal $\Lambda$ mass in the plane of the effective masses
$m({\rm p} \pi^-)$ $\vs$ $m({\rm \overline{p}} \pi^+)$, shown in
Figures~\ref{mass}c and~\ref{mass}d. The remaining $\koskos$
contamination is estimated by Monte Carlo simulation to be less than
1\%. The normalisation of the $\koskos$ Monte Carlo background is
determined from our previous measurement of this channel
\cite{b12}. Within the available statistics, the hypothesis of an
isotropic distribution of the $\Lambda\llbar$ signal is verified.

\subsection{$\mathbf{\Sigma^0}$ identification}

\begin{figure}
\begin{center}
  \epsfig{file=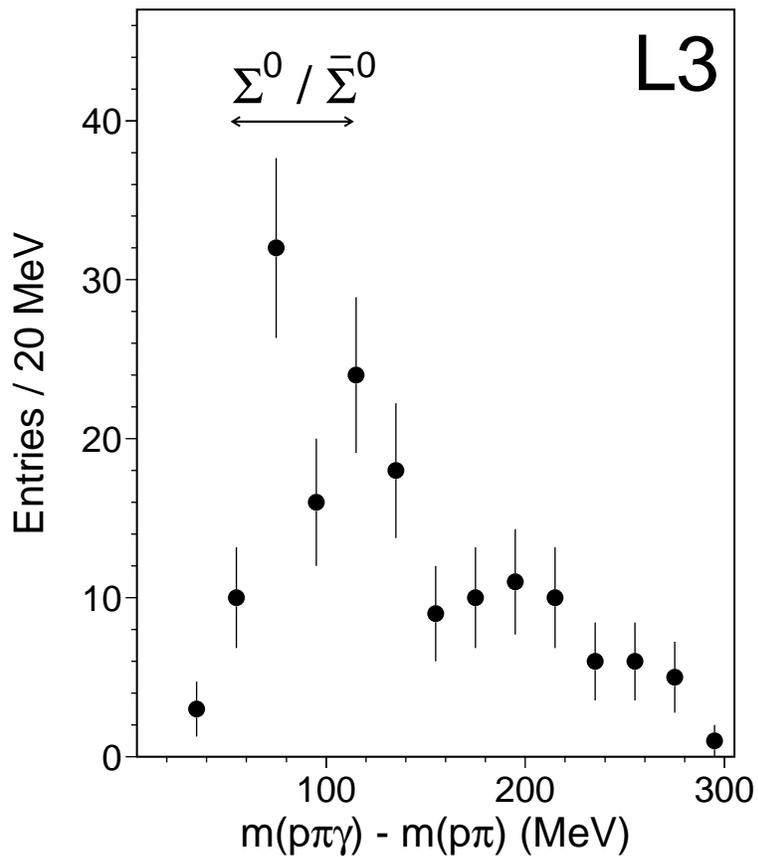,width=10cm}
\end{center} 
\caption{ Distribution of the mass difference between the $\Lambda\gamma$ 
 ($\mathrm{p}\pi\gamma$) and the $\Lambda$ ($\mathrm{p}\pi$)
systems. All possible combinations of a photon and a $\Lambda$ or
$\llbar$ candidate are shown.}
\label{sigmass}
\end{figure}

The reconstruction of $\Sigma^0$ and $\ssbar$ candidates is performed
by combining the selected $\Lambda$ and $\llbar$ with photon
candidates. A photon candidate is defined as a shower in the
electromagnetic calorimeter with at least two adjacent crystals and an
energy between 50 MeV and 200 MeV. Monte Carlo studies show that 91\%
of the photons emitted by a $\Sigma^0$ have an energy below 200 MeV,
which is compatible with the nominal $\Sigma^0$ mass of 1.193
GeV. Photon isolation criteria are also applied; there must be no
charged tracks within 200 mrad around the photon direction and the
cosine of the angle between the antiproton and the photon directions
must be less than 0.8. To identify the $\Sigma^0$, the mass difference
\mbox{$ \Delta m = m({\rm p} \pi \gamma) - m({\rm p} \pi)$} is used,
as presented in Figure~\ref{sigmass}. A $\Sigma^0$ candidate
corresponds to the mass interval \mbox{47 MeV} $< \Delta m <$
\mbox{107 MeV}. Out of the 66 selected $\Lambda\llbar$ events, 31 have
$\Sigma^0$ candidates.

\section{Exclusive $\mathbf{\lalabar}$ and $\mathbf{\sisibar}$ identification}

\begin{figure}
\begin{center}
\includegraphics[width=15cm]{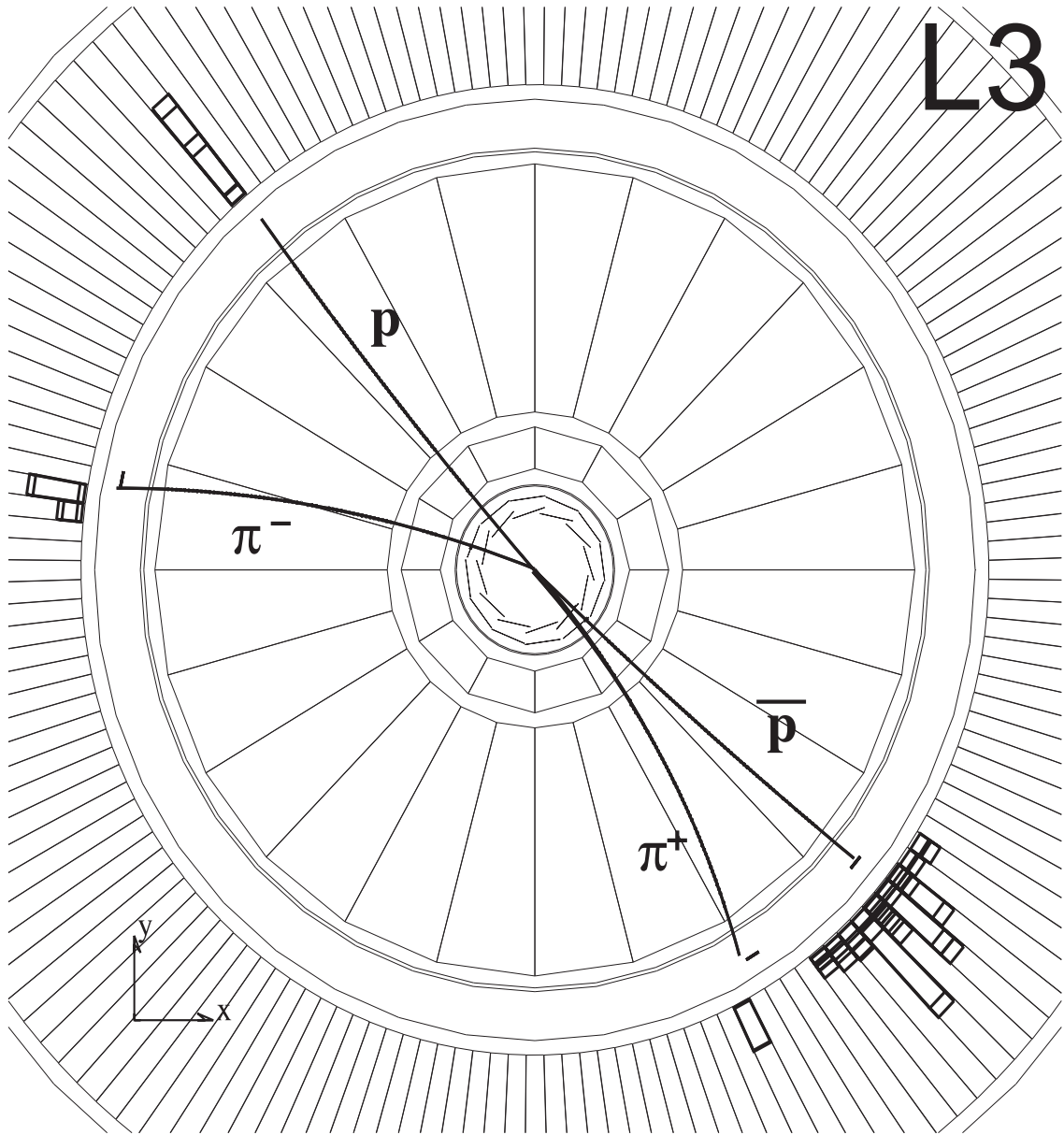}
\end{center}
\caption {A typical $\gamma \gamma \ra \Lambda \llbar$ event, displayed in 
the transverse plane. It illustrates the higher momentum of the proton
and antiproton in the $\Lambda$ and $\llbar$ decays and the separation
in the electromagnetic calorimeter between the large antiproton signal
and the small energy deposit of pions and protons.}
\label{event}
\end{figure}

  In order to select the events from the exclusive reactions $\gamma
  \gamma \ra \lalabar$ and $\gamma \gamma \ra \sisibar$, the
  transverse momentum of the four charged particles, $P_T$, is
  required to be less than 0.5 GeV. This cut rejects events containing
  contributions from other final states such as $\Xi^0
  \overline{\Xi}^0$ or $\Sigma^0(1385) \ssbar(1385)$. Some of these
  states can still pass the $ P_T$ cut, but their contribution to the
  final sample is negligible, given the magnitude of their cross
  sections \cite{b3}. The $P_T$ requirement rejects less than 8\% of
  events corresponding to the exclusive final states $\Lambda \llbar$
  and $\Sigma^0 \ssbar$. Since the photons emitted by the $\Sigma^0$
  candidates have a low energy, they give a small contribution to the
  total transverse momentum imbalance. The final sample contains 33
  events. The numbers of selected events for the different $\rm
  e^+e^-$ centre-of-mass energies are listed in Table~\ref{tab}. A
  typical $\Lambda \llbar$ event is shown in Figure~\ref{event}.

  The relative proportions of $\lalabar$ and $\sisibar$ final states
in the sample are determined as follows. The event is labelled as
$\sisibar$-like if a $\Sigma^0$ or a $\ssbar$ candidate is observed
and as $\lalabar$-like otherwise.  With these criteria, 19 $\Lambda
\llbar$-like and 14 $\Sigma^0 \ssbar$-like events are found in the
data. The true fractions $r_j$ ($j=\lalabar,\sisibar)$ of the two
components are determined by a maximum extended likelihood fit with
the constraint $r_{\lalabar}+r_{\sisibar}=1$. The likelihood function
to be maximized is:

$${\cal L} = \frac{n_t^{N_t}~e^{-n_t}}{N_t!}  \prod_{i}
\frac{n_i^{N_i}~e^{-n_i}} {N_i!} \; ,$$

\noindent
where $ N_t$ and $ n_t$ correspond respectively to the total number of
observed and expected events, and $N_i$ and $n_i$ to the number of
observed and expected $i$-like events. The latter is given by:

$$ n_i = ( \sum_j p_{ij} r_j ) n_t \;,$$

\noindent
where $p_{ij}$ is the probability of identifying an event
corresponding to the final state $j$ as $i$-like. The relative
probabilities $p_{ij}$ are determined by Monte Carlo and shown in
Table~\ref{pij} together with their statistical uncertainties. The
fractions $r_j$ and the number of events $n_t r_j$ obtained by the fit
are given in Table~\ref{pij2}.

The cross sections for the $\gamma\gamma\ra\Lambda \ssbar$ and
$\gamma\gamma\ra\Sigma^0 \llbar$ processes are predicted to be
negligible compared to the other channels~\cite{b3}. In order to test
this assumption, also an analysis with the three components
$\lalabar$, $\Lambda\ssbar+\Sigma^0\llbar$ and $\sisibar$ is carried
out. The $\Lambda\ssbar+\Sigma^0\llbar$ fraction is measured to be
compatible with zero within a large uncertainty.

\section{Results}

   The production cross sections $\sigma (\epem \ra \epem \lalabar)$
and $\sigma (\epem \ra \epem \sisibar)$ are measured as a function of
the centre-of-mass energy. They refer to the following phase-space
cuts: the effective mass of the $\Lambda \llbar$ pair, $m_{\lalabar}$,
less than 3.5 GeV, $| \cos \theta^* | < 0.6$ and $P_T < 0.5$ GeV. In
the cross section determination it is assumed that the fractions $r_i$
are independent of $\rts$. The results are summarised in
Table~\ref{tab2}.

  The detection efficiency is determined by Monte Carlo for each data
  taking period. It takes into account the $\rm \Lambda \ra p \pi$
  branching ratio and track geometrical acceptance ($\simeq$ 6\%), the
  baryon identification criteria ($\simeq$ 26\%) and the track trigger
  efficiency ($\simeq$ 10\%). The efficiency of higher level triggers
  ($\simeq$ 90\%) is estimated from the data themselves, using
  prescaled events. The contribution of the different selection cuts
  to the detection efficiency is detailed in Table~\ref{deteff}. The
  total efficiencies for each data set are listed in Table~\ref{tab}.

  The dominant source of systematic uncertainty is the selection
  procedure (7\%); other sources are the finite Monte Carlo statistics
  (5\%) and the determination of the trigger efficiency (3\%). The
  Monte Carlo contribution includes the uncertainty on the $p_{ij}$
  probabilities used in the determination of the fractions $r_i$.

   The cross sections $\sigma( \gamma \gamma \ra \Lambda \llbar)$ and
$\sigma(\gamma \gamma \ra \Sigma^0 \ssbar)$ in real photon collisions
are extracted as a function of $W_{\gamma \gamma}$ by deconvoluting
the two-photon luminosity function and the form
factor~\cite{b14}. They are presented in Table~\ref{ggresults2}. For
the $\gamma \gamma \ra \Sigma^0 \ssbar$ case, the number of selected
events as a function of $W_{\gamma\gamma}$ is obtained from the
corresponding $m_{\Lambda \llbar}$ distribution, within a 4.0\%
uncertainty. The efficiencies and luminosity functions are evaluated
for each $W_{\gamma \gamma}$ interval and centre-of-mass energy. The
efficiencies increase with $W_{\gamma\gamma}$ reflecting the expected
rise in the detector acceptance. The trigger and track identification
efficiencies do not depend on $W_{\gamma\gamma}$. An additional
systematic uncertainty of 5\%, due to the choice of the photon form
factor, is included.

\begin{figure}
\begin{center}
\epsfig{file=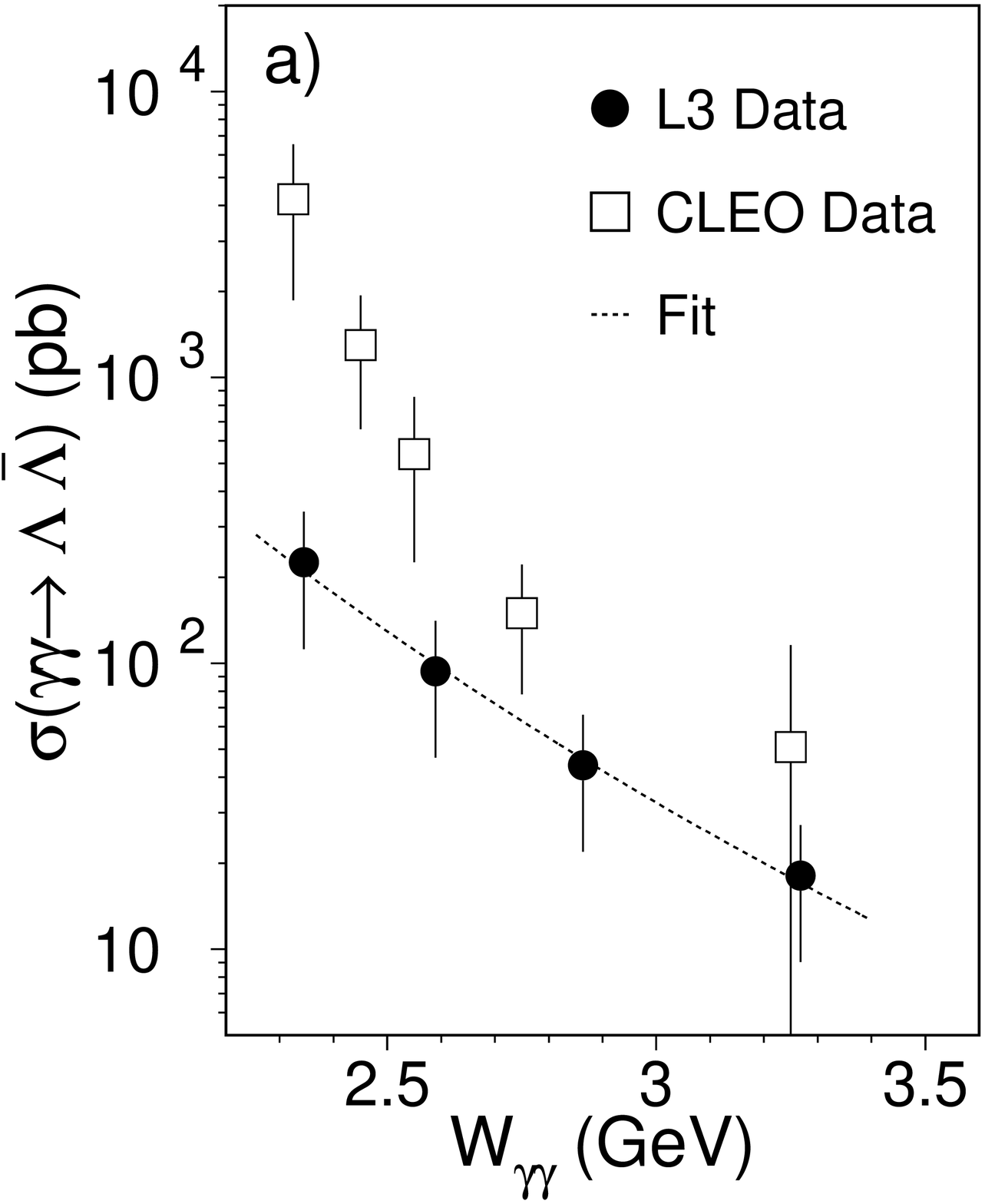,height=8.5cm} \hspace{0.5cm}
\epsfig{file=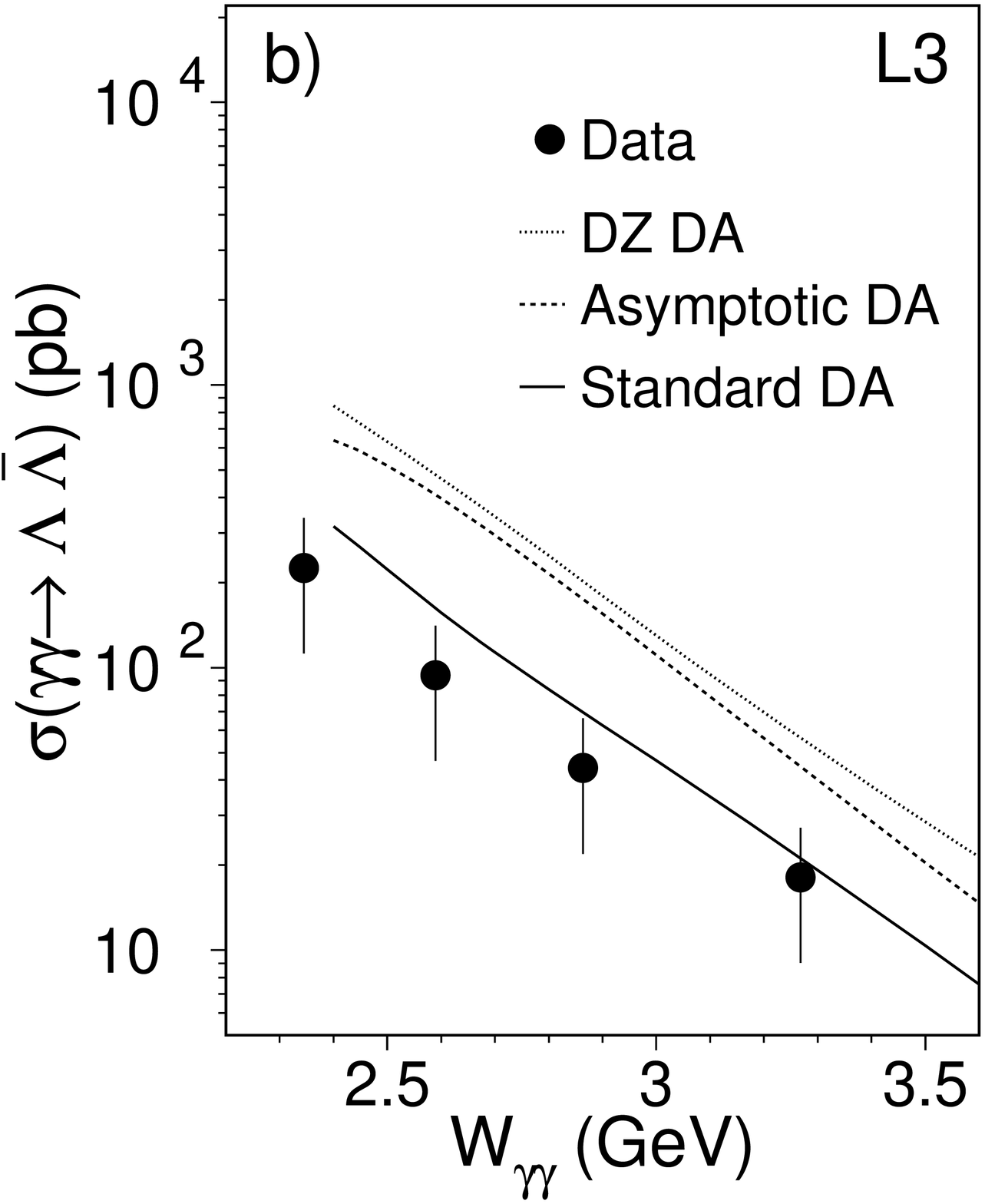,height=8.5cm} \vspace{1cm}\\
\epsfig{file=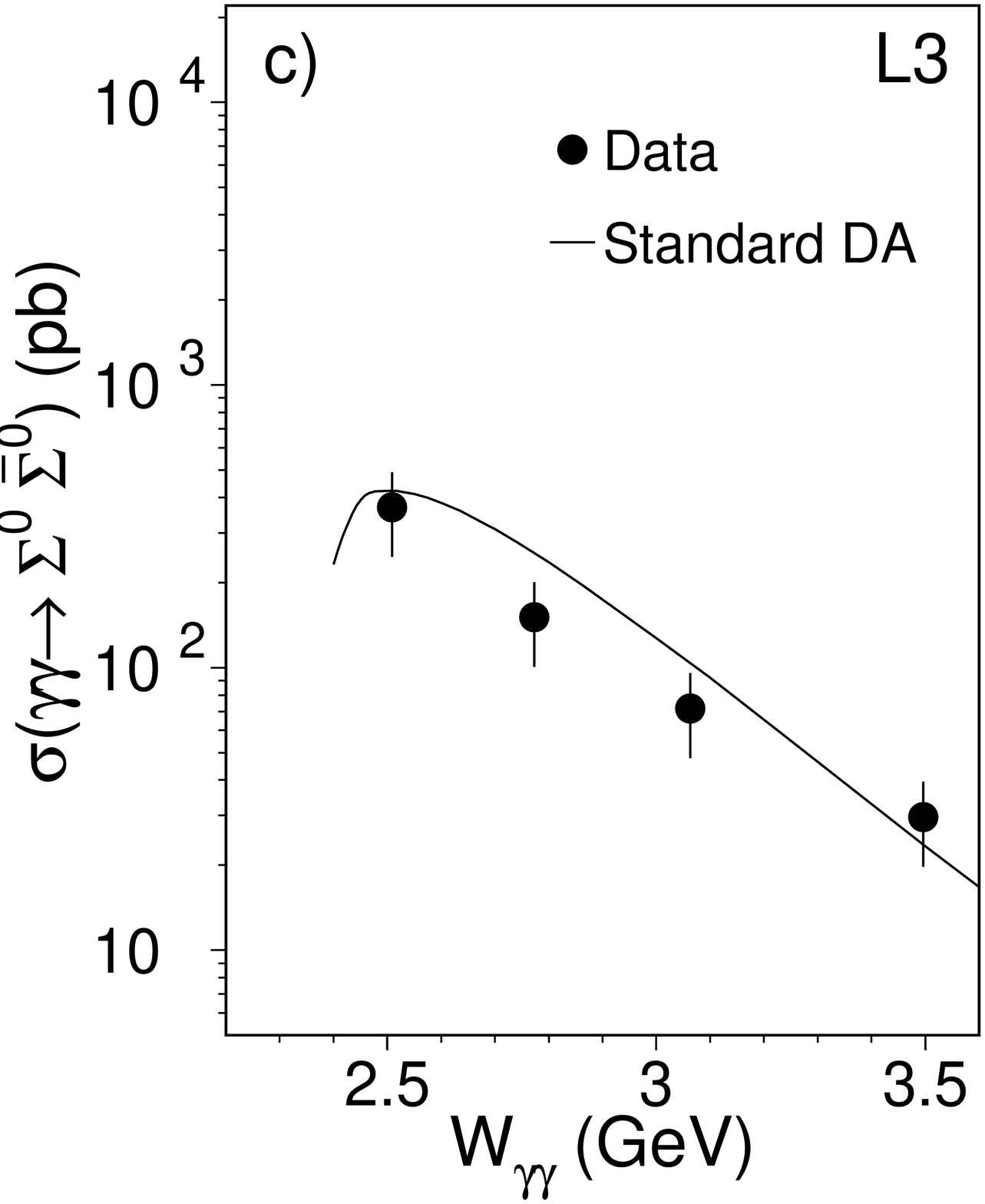,height=8.5cm}
\end{center}
\caption{Measurements of the $\gamma \gamma \ra \lalabar$ and $\gamma\gamma \ra \sisibar$ cross sections as a function of $W_{\gamma\gamma}$. In a) the $\gamma \gamma \ra \lalabar$ cross section is compared to the one obtained by CLEO $\lbrack 4 \rbrack$. The dashed line shows the power law fit described in the text. In b) and c) the $\sigma(\gamma \gamma \ra \Lambda \llbar)$ and $\sigma(\gamma \gamma \ra \Sigma^0 \ssbar)$ measurements are compared to the calculations of Reference {\protect \citen{b3}}. Statistical and systematic uncertainties are added in quadrature.}
\label{comp}
\end{figure}

   Figure~\ref{comp}a compares the present $\sigma(\gamma \gamma \ra
\lalabar)$ measurement with that of CLEO. The mass dependence of CLEO is steeper than the one we observe. Our data,
fitted with a function of the form $\sigma \propto \; W^{-n} $, gives
a value n = $ 7.6 \pm 3.9$. The quark-diquark model predicts n=6, and
a three quark model n=10 \cite{b30}. In Figures~\ref{comp}b
and~\ref{comp}c, the $\gamma \gamma \ra \Lambda \llbar$ and $\gamma
\gamma \ra \Sigma^0 \ssbar$ cross section measurements are compared to
the predictions of recent quark-diquark model calculations
\cite{b3}. This model considers three different distribution
amplitudes (DA) for the diquarks. The absolute predictions using the
standard distribution amplitude (Standard DA) reproduce well our data.
The asymptotic DA \cite{b32} and DZ-DA \cite{b31} models are
excluded.

\subsection*{Acknowledgments}

    We thank C. F. Berger and W. Schweiger for very useful discussions
and for providing us their theoretical predictions.

%
\newpage
\section*{Author List}
\input namelist251.tex \newpage

%
\newpage 
\bibliographystyle{l3style}


\newpage

\begin{table}[H]
\begin{center}
\begin{tabular}{|c|c|c|c|c|} \hline 
$\sqrt{s}$ (GeV) & Luminosity (pb$^{-1}$) & Efficiency (\%) & Events
\\ \hline 91 & 157 & $2.38 \pm 0.11 $ & \phantom{0}8 \\ 161$-$172 &
\phantom{0}20 & $1.72 \pm 0.07 $ & \phantom{0}3 \\ 183 & \phantom{0}52
& $1.97 \pm 0.09 $ & \phantom{0}1 \\ 189 & 172 & $1.78 \pm 0.07 $ &
\phantom{0}3 \\ 192$-$202 & 230 & $1.94 \pm 0.07 $ & 10 \\ 205$-$208 &
213 & $1.75 \pm 0.07 $ & \phantom{0}8 \\ \hline
\end{tabular} 

\caption{ Integrated luminosity, overall efficiency and number of selected $\epem \ra \epem \lalabar$ and $\epem \ra \epem \sisibar$ events for each data taking period. The efficiency refers to the phase-space cuts: $ 2.23 <m_{\Lambda \llbar} < 3.5$ GeV, $|\cos \theta^*|<0.6$ and $P_T < 0.5$ GeV. The quoted uncertainties are statistical.}
\label{tab}
\end{center}
\end{table}

\begin{table}[H]
\begin{center}
\begin{tabular}{|c | c |c c|} \hline 
Identified as & $N_i$ & \multicolumn{2}{c|}{Selection probability $
p_{ij}$ (\%)} \\ \vspace{-0.2cm} & & &\\ & & Generated as $\Lambda
\overline{\Lambda}$ & Generated as $\Sigma^0 \ssbar$ \\\hline $\Lambda
\overline{\Lambda}$-like & 19 & 88.0 $\pm$ 0.8 & 39.1 $\pm$ 1.0 \\
$\Sigma^0 \ssbar $-like& 14 & 12.0 $\pm$ 0.8 & 60.9 $\pm$ 1.0 \\
\hline
\end{tabular}

\caption{Numbers of observed events $N_i$ identified as $\lalabar$-like and
$\sisibar$-like and relative probabilities $p_{ij}$ of identifying an
event generated in the final state $j$ as $i$-like.}
\label{pij}
\end{center}
\end{table}

\begin{table}[H]
\begin{center}
\begin{tabular}{|c | c | c |} \hline
Final state & Fraction $ r_j$ & Events $n_t r_j$ \\\hline $\Lambda
\overline{\Lambda}$ & $0.38 \pm 0.18$ & $12.5 \pm 6.1$ \\
$\Sigma^0 \ssbar $ & $0.62 \pm 0.18$ & $20.5 \pm 6.5$ \\ \hline
\end{tabular}

\caption{Results of the fit for the fractions $r_j$ and the number $n_t r_j$ of $\lalabar$ and $\sisibar$ final states.}
\label{pij2}
\end{center}
\end{table}

\begin{table}[H]
\begin{center}
\begin{tabular}{|c|c|c|} \hline 
$ \sqrt{s}$ (GeV) & $\sigma(\epem \ra \epem \Lambda \llbar) $ (pb)&
$\sigma(\epem \ra \epem \Sigma^0 \ssbar)$ (pb) \\ \hline 91 & $0.81
\pm 0.48 \pm 0.07 $& $1.33\pm 0.61 \pm 0.12$ \\
%
%
 161$-$208 & $0.75 \pm 0.39 \pm 0.07$ & $1.23 \pm 0.43 \pm 0.11$ \\
 \hline
\end{tabular} 

\caption{ The $\epem \ra \epem \Lambda \llbar$ and $\epem \ra \epem \Sigma^0 \ssbar$ cross sections for $ 2.23 <m_{\Lambda \llbar} < 3.5$ GeV, $|\cos \theta^*|<0.6$ and $ P_T < 0.5$ GeV. The first uncertainty is statistical, the second systematic. }
\label{tab2}
\end{center}
\end{table}

\begin{table}[H]
\begin{center}
\begin{tabular}{|l|c|} \hline 
\multicolumn{1}{|c|}{ Cut} & Acceptance (\%) \\ \hline Four tracks &
10 \\ Two secondary vertices & 55 \\ Photon conversion rejection & 93
\\ $dE/dx$ compatibility & 98 \\ $dE/dx$ or $E_T/p_T$ & 77 \\
$\koskos$ rejection & 92 \\ $|\cos \theta^*|$ & 83 \\ $\Lambda$ vs
$\llbar$ mass & 94 \\ $P_T$ & 92 \\\hline Total & 2.5 \\ \hline
\end{tabular}

\caption{Average acceptance of the different selection cuts used in the analysis. The overall efficiency takes also into account the acceptance of the trigger system ($\simeq 80\%$).}
\label{deteff}
\end{center}
\end{table}

\begin{table}[H]
\begin{center}
\begin{tabular}{|c|c|c||c|c|c|} \hline 

$ W_{\gamma \gamma}$ (GeV) & $\langle W_{\gamma \gamma} \rangle$ (GeV)
& $\sigma(\gamma \gamma \ra \lalabar)$ (pb)& $ W_{\gamma \gamma}$
(GeV) & $\langle W_{\gamma \gamma} \rangle$ (GeV) & $\sigma(\gamma
\gamma \ra \sisibar) (\rm pb)$ \\ \hline
%
2.2 $-$ 2.5 & 2.34 & $226 \pm 111 \pm \phantom{0}23$ &
2.4 $-$ 2.7 & 2.51 & $369 \pm 116 \pm \phantom{0}41$ \\  
2.5 $-$ 2.7 & 2.59 & $\phantom{0}92 \pm \phantom{0}45 \pm \phantom{0}10$ &
2.7 $-$ 2.9 & 2.77 & $151 \pm \phantom{0}48 \pm \phantom{0}17$ \\  
2.7 $-$ 3.1 & 2.86 & $\phantom{0}44 \pm \phantom{0}22 \pm \phantom{00}5$ &
2.9 $-$ 3.3 & 3.06 & $\phantom{0}72 \pm \phantom{0}23 \pm \phantom{00}8$ \\  
3.1 $-$ 3.5 & 3.27 & $\phantom{0}18 \pm \phantom{00}9 \pm \phantom{00}2$ &
3.3 $-$ 3.8 & 3.50 & $\phantom{0}30 \pm \phantom{00}9 \pm \phantom{00}3$ \\ 
\hline
\end{tabular}

\caption{ The $\gamma \gamma \ra \lalabar$ and  $\gamma \gamma \ra \Sigma^0 \ssbar$ cross sections as a function of $ W_{\gamma \gamma}$ for $| \cos \theta^* | < 0.6$ and $ P_T < 0.5$ GeV. The central value $\langle  W_{\gamma \gamma} \rangle $ of each bin corresponds to an average according to a $W^{-8}$ distribution. The first uncertainty is statistical, the second systematic. }
\label{ggresults2}
\end{center}
\end{table}

\newpage

\newpage

\newpage

\newpage

\newpage

\end{document}

%% file: namelist251.tex
\typeout{   }     
\typeout{Using author list for paper 251 -- ? }
\typeout{$Modified: Jul 15 2001 by smele $}
\typeout{!!!!  This should only be used with document option a4p!!!!}
\typeout{   }
%
%
%
%
%
%

\newcount\tutecount  \tutecount=0
\def\tutenum#1{\global\advance\tutecount by 1 \xdef#1{\the\tutecount}}
\def\tute#1{$^{#1}$}
\tutenum\aachen            
\tutenum\nikhef            
\tutenum\mich              
\tutenum\lapp              
\tutenum\basel             
\tutenum\lsu               
\tutenum\beijing           
\tutenum\berlin            
\tutenum\bologna           
\tutenum\tata              
\tutenum\ne                
\tutenum\bucharest         
\tutenum\budapest          
\tutenum\mit               
\tutenum\panjab            
\tutenum\debrecen          
\tutenum\florence          
\tutenum\cern              
\tutenum\wl                
\tutenum\geneva            
\tutenum\hefei             
\tutenum\lausanne          
\tutenum\lyon              
\tutenum\madrid            
\tutenum\florida           
\tutenum\milan             
\tutenum\moscow            
\tutenum\naples            
\tutenum\cyprus            
\tutenum\nymegen           
\tutenum\caltech           
\tutenum\perugia           
\tutenum\peters            
\tutenum\cmu               
\tutenum\potenza           
\tutenum\prince            
\tutenum\riverside         
\tutenum\rome              
\tutenum\salerno           
\tutenum\ucsd              
\tutenum\sofia             
\tutenum\korea             
\tutenum\purdue            
\tutenum\psinst            
\tutenum\zeuthen           
\tutenum\eth               
\tutenum\hamburg           
\tutenum\taiwan            
\tutenum\tsinghua          

{
\parskip=0pt
\noindent
{\bf The L3 Collaboration:}
\ifx\selectfont\undefined
 \baselineskip=10.8pt
 \baselineskip\baselinestretch\baselineskip
 \normalbaselineskip\baselineskip
 \ixpt
\else
 \fontsize{9}{10.8pt}\selectfont
\fi
\medskip
\tolerance=10000
\hbadness=5000
\raggedright
\hsize=162truemm\hoffset=0mm
\def\r{\rlap,}
\noindent

P.Achard\r\tute\geneva\ 
O.Adriani\r\tute{\florence}\ 
M.Aguilar-Benitez\r\tute\madrid\ 
J.Alcaraz\r\tute{\madrid,\cern}\ 
G.Alemanni\r\tute\lausanne\
J.Allaby\r\tute\cern\
A.Aloisio\r\tute\naples\ 
M.G.Alviggi\r\tute\naples\
H.Anderhub\r\tute\eth\ 
V.P.Andreev\r\tute{\lsu,\peters}\
% ...
\newpage
\begin{list}{A}{\itemsep=0pt plus 0pt minus 0pt\parsep=0pt plus 0pt minus 0pt
                \topsep=0pt plus 0pt minus 0pt}
\item[\aachen]
 I. Physikalisches Institut, RWTH, D-52056 Aachen, FRG$^{\S}$\\
 III. Physikalisches Institut, RWTH, D-52056 Aachen, FRG$^{\S}$
\item[\nikhef] National Institute for High Energy Physics, NIKHEF, 
     and University of Amsterdam, NL-1009 DB Amsterdam, The Netherlands
\item[\mich] University of Michigan, Ann Arbor, MI 48109, USA
\item[\lapp] Laboratoire d'Annecy-le-Vieux de Physique des Particules, 
     LAPP,IN2P3-CNRS, BP 110, F-74941 Annecy-le-Vieux CEDEX, France
\item[\basel] Institute of Physics, University of Basel, CH-4056 Basel,
     Switzerland
\item[\lsu] Louisiana State University, Baton Rouge, LA 70803, USA
\item[\beijing] Institute of High Energy Physics, IHEP, 
  100039 Beijing, China$^{\triangle}$ 
\item[\berlin] Humboldt University, D-10099 Berlin, FRG$^{\S}$
\item[\bologna] University of Bologna and INFN-Sezione di Bologna, 
     I-40126 Bologna, Italy
\item[\tata] Tata Institute of Fundamental Research, Mumbai (Bombay) 400 005, India
\item[\ne] Northeastern University, Boston, MA 02115, USA
\item[\bucharest] Institute of Atomic Physics and University of Bucharest,
     R-76900 Bucharest, Romania
\item[\budapest] Central Research Institute for Physics of the 
     Hungarian Academy of Sciences, H-1525 Budapest 114, Hungary$^{\ddag}$
\item[\mit] Massachusetts Institute of Technology, Cambridge, MA 02139, USA
\item[\panjab] Panjab University, Chandigarh 160 014, India.
\item[\debrecen] KLTE-ATOMKI, H-4010 Debrecen, Hungary$^\P$
\item[\florence] INFN Sezione di Firenze and University of Florence, 
     I-50125 Florence, Italy
\item[\cern] European Laboratory for Particle Physics, CERN, 
     CH-1211 Geneva 23, Switzerland
\item[\wl] World Laboratory, FBLJA  Project, CH-1211 Geneva 23, Switzerland
\item[\geneva] University of Geneva, CH-1211 Geneva 4, Switzerland
\item[\hefei] Chinese University of Science and Technology, USTC,
      Hefei, Anhui 230 029, China$^{\triangle}$
\item[\lausanne] University of Lausanne, CH-1015 Lausanne, Switzerland
\item[\lyon] Institut de Physique Nucl\'eaire de Lyon, 
     IN2P3-CNRS,Universit\'e Claude Bernard, 
     F-69622 Villeurbanne, France
\item[\madrid] Centro de Investigaciones Energ{\'e}ticas, 
     Medioambientales y Tecnol\'ogicas, CIEMAT, E-28040 Madrid,
     Spain${\flat}$ 
\item[\florida] Florida Institute of Technology, Melbourne, FL 32901, USA
\item[\milan] INFN-Sezione di Milano, I-20133 Milan, Italy
\item[\moscow] Institute of Theoretical and Experimental Physics, ITEP, 
     Moscow, Russia
\item[\naples] INFN-Sezione di Napoli and University of Naples, 
     I-80125 Naples, Italy
\item[\cyprus] Department of Physics, University of Cyprus,
     Nicosia, Cyprus
\item[\nymegen] University of Nijmegen and NIKHEF, 
     NL-6525 ED Nijmegen, The Netherlands
\item[\caltech] California Institute of Technology, Pasadena, CA 91125, USA
\item[\perugia] INFN-Sezione di Perugia and Universit\`a Degli 
     Studi di Perugia, I-06100 Perugia, Italy   
\item[\peters] Nuclear Physics Institute, St. Petersburg, Russia
\item[\cmu] Carnegie Mellon University, Pittsburgh, PA 15213, USA
\item[\potenza] INFN-Sezione di Napoli and University of Potenza, 
     I-85100 Potenza, Italy
\item[\prince] Princeton University, Princeton, NJ 08544, USA
\item[\riverside] University of Californa, Riverside, CA 92521, USA
\item[\rome] INFN-Sezione di Roma and University of Rome, ``La Sapienza",
     I-00185 Rome, Italy
\item[\salerno] University and INFN, Salerno, I-84100 Salerno, Italy
\item[\ucsd] University of California, San Diego, CA 92093, USA
\item[\sofia] Bulgarian Academy of Sciences, Central Lab.~of 
     Mechatronics and Instrumentation, BU-1113 Sofia, Bulgaria
\item[\korea]  The Center for High Energy Physics, 
     Kyungpook National University, 702-701 Taegu, Republic of Korea
\item[\purdue] Purdue University, West Lafayette, IN 47907, USA
\item[\psinst] Paul Scherrer Institut, PSI, CH-5232 Villigen, Switzerland
\item[\zeuthen] DESY, D-15738 Zeuthen, 
     FRG
\item[\eth] Eidgen\"ossische Technische Hochschule, ETH Z\"urich,
     CH-8093 Z\"urich, Switzerland
\item[\hamburg] University of Hamburg, D-22761 Hamburg, FRG
\item[\taiwan] National Central University, Chung-Li, Taiwan, China
\item[\tsinghua] Department of Physics, National Tsing Hua University,
      Taiwan, China
\item[\S]  Supported by the German Bundesministerium 
        f\"ur Bildung, Wissenschaft, Forschung und Technologie
\item[\ddag] Supported by the Hungarian OTKA fund under contract
numbers T019181, F023259 and T024011.
\item[\P] Also supported by the Hungarian OTKA fund under contract
  number T026178.
\item[$\flat$] Supported also by the Comisi\'on Interministerial de Ciencia y 
        Tecnolog{\'\i}a.
\item[$\sharp$] Also supported by CONICET and Universidad Nacional de La Plata,
        CC 67, 1900 La Plata, Argentina.
\item[$\triangle$] Supported by the National Natural Science
  Foundation of China.
\end{list}
}
\vfill
